\documentclass[11pt]{article}

\usepackage{amsmath}
\usepackage{graphicx}
\usepackage{indentfirst}
\usepackage{amssymb}
\usepackage{cite}
\usepackage{color}
\usepackage{subfigure}
\usepackage{xcolor}
\usepackage[breaklinks=true]{hyperref}

\usepackage{mathrsfs}
\usepackage{amsmath}

\begin{document}

\title{Holographic entropy bound and a special class of spherical systems in cosmology
}
\date{}
\maketitle

\begin{center}
	\author{Hao Yu$~^{a,}$\footnote{yuhaocd@cqu.edu.cn},
		Zichao Lin$^{a,b,}$\footnote{linzch22@hust.edu.cn},
		Jin Li~$^{a,}$\footnote{cqujinli1983@cqu.edu.cn (corresponding author)}}
\end{center}

\begin{center}
	$^a$ Physics Department, Chongqing University, Chongqing 401331, China\\
	$^b$  School of Physics, Huazhong University of Science and Technology, Wuhan, Hubei 430074, China
\end{center}

\begin{abstract}
The holographic entropy bound is discussed in cosmology. Inspired by the work of Fischler and Susskind [hep-th/9806039], we aim to define a special class of spherical systems in cosmology, within which the entropy of matter remains compliant with the holographic entropy bound throughout the evolution of the universe, irrespective of the universe's components. It is found that  if the entropy of matter per unit co-moving volume is bounded from above, such a special class of spherical systems indeed exists. Moreover, the matter contained within a unit co-moving volume can be replaced by a black hole of the same mass-energy. Provided that the entropy of the black hole consistently exceeds that of the matter it replaces, there is also a unified definition for these special spherical systems.
\end{abstract}

\maketitle

\section{Introduction}\label{sec1}

The first entropy bound, i.e., the Bekenstein bound~\cite{Bekenstein:1980jp}, was proposed by Jacob Bekenstein in 1981, which points out that the total entropy in a spherical system of radius $R$ should follow an inequality $S\leq  \frac{2\pi k_B E R}{\hbar\, c}$, where $E$ is the total mass-energy contained within the system. More than a decade later, the holographic principle~\cite{tHooft:1993dmi,Stephens:1993an,Susskind:1994vu,Witten:1998qj,Susskind:1998dq,Bousso:2002ju} indicates that the description of a volume of space is encoded on its lower-dimensional boundary. Thus, the degree of freedom of a spatial region resides on its boundary and the maximum entropy of the spatial region should be its area in Planck units, which is also known as the holographic (entropy) bound. Since the proposal of these two entropy bounds, physicists have suggested various entropy bounds~\cite{Susskind:1994vu,Bousso:1999xy,Flanagan:1999jp,Brustein:1999md,Verlinde:2000wg}. For a review, refer to Refs.~\cite{Wald:1999vt,Bousso:2002ju}.

Most entropy bounds, particularly those arising from black hole physics, are inherently rooted in static space-times. Consequently, they are prone to violations in dynamical  systems~\cite{Bousso:2002ju}, such as collapsing stars~\cite{Bousso:1999xy,Easther:1999gk}, the universe~\cite{Susskind:1998dq}, and weakly gravitating spherical thermodynamic systems in asymptotically flat space~\cite{Bousso:2002ju}. To address the problem and preserve the holographic principle, Bousso introduced a covariant version of the entropy bound, i.e., the Bousso bound~\cite{Bousso:1999xy,Flanagan:1999jp,Strominger:2003br}, which states that the entropy should be evaluated on null hypersurfaces. The Bousso bound can be applicable to arbitrarily curved space-time~\cite{Bousso:2003kb,Gao:2004mc,Pesci:2007rp,Pesci:2008yy}, so it is considered a potential imprint of quantum gravity. 
	
In 1998, Fischler and Susskind proposed to apply the holographic bound from nearly flat space to both flat and non-flat universes~\cite{Fischler:1998st}. They found that, for the entropy contained within a volume of coordinate size $R_H$, the practicability of the holographic bound depends on multiple parameters in cosmology, so it does not universally hold true. In fact, as early as 1989, Bekenstein had already discovered that his entropy bound is violated in the early universe when applied to the entropy contained within the particle horizon of a given observer. Surprisingly, this phenomenon could potentially help solve the cosmological singularity problem~\cite{Bekenstein:1989wf,Powell:2020nzc}. In a similar work~\cite{Yu:2023dze}, it was also found that, as the universe evolves, the entropy contained within a unit co-moving volume can exceed both the Bekenstein bound and the holographic bound. These studies indicate that applying directly entropy bounds in cosmological contexts faces significant challenges.

\section{A question on entropy bounds in cosmology} 

When entropy bounds are applied to cosmology, we can refer to them as cosmological entropy bounds. If entropy bounds seek to determine the maximum entropy contained within a closed spherical system, then cosmological entropy bounds can be seen as extending the concept for a closed system to an open  system (a spherical system with an evolving boundary). However, the majority of existing studies~\cite{Bekenstein:1989wf,Fischler:1998st,Bousso:2002ju,Brustein:2007hd,Yu:2023dze,Powell:2020nzc} demonstrate that cosmological entropy bounds can be readily violated in a general spherical system. An analysis of previous studies reveals that the primary reason for the violation of entropy bounds in cosmology lies in the choice of the spherical system. In these studies, the radii of spherical systems are typically the characteristic length scales in cosmology, such as the particle horizon of a given observer ~\cite{Bekenstein:1989wf,Powell:2020nzc}, the cosmological apparent horizon~\cite{Bak:1999hd}, the Hubble horizon~\cite{Easther:1999gk,Veneziano:1999ts,Brustein:1999ua}, the radius of $R\sim H^{-1}\sim t$~\cite{Kaloper:1999tt}, and so forth. For these spherical systems, their internal entropy can readily violate entropy bounds at certain cosmological epochs. To elucidate the phenomenon, we perform the following semi-quantitative analysis. Consider a spherical system, within which the maximum entropy permitted by a cosmological entropy bound takes the form: $S_{max}=f(t)\, R^n$ ($n$ remains constant in most cases, such as $n=2$ for the holographic bound, and $n=4$ for the Bekenstein bound). Here, $R$ is the radius of the spherical system and $f(t)$ is a variable independent of $R$. In our analysis, we restrict consideration to extensive entropy formulations, thereby excluding non-extensive entropy measures, such as Tsallis entropy~\cite{Tsallis:1987eu,Wilk:1999dr,Tsallis:2012js}. Thus, the entropy inside the spherical system can be expressed as $S=s(t) R^3$ with $s(t)$ being a time-dependent entropy density. In an expanding universe, since the radius $R$ evolves with the scale factor, it must be vanishingly small in the early universe but grow to an extremely large scale at late times. As a result, entropy bounds with $n<3$ are readily violated in the late universe due to the increase of $R$. Similarly, entropy bounds with $n>3$ are violated in the early universe provided $R$ is sufficiently small. For the case of $n=3$, to guarantee the validity of entropy bounds during the entire evolution of the universe, the coefficient $f(t)$ must remain strictly greater than the entropy density $s(t)$ at all cosmological epochs. However, the implementation of this condition is  challenging, since the entropy density $s(t)$ is highly sensitive to both the evolution and components of the universe. 

Based on the analysis, an intriguing question arises: Does there exist a special class of spherical systems in cosmology, within which the entropy of matter remains compliant with a given entropy bound throughout the evolution of the universe, irrespective of the universe's components? Next, we employ the holographic bound as a case to answer this question. Remarkably, when the components of the universe are properly specified, such a special class of spherical systems can usually be found for the holographic bound. For example, in a universe full of radiation, we consider a spherical system of radius $R$. The holographic principle conjectures that the entropy of matter inside a given spatial region can not exceed the area of its boundary surface:
\begin{eqnarray}\label{1}
		S\leq\frac{k_B A }{4l_p^2},
\end{eqnarray}
where $A$ is the boundary area.	Therefore, the evolution of $R$ need satisfy 
\begin{eqnarray}
\frac{\frac43\pi R^3}{a(t)^3V_0} S_0\leq \frac{4 \pi R^2k_B}{4\,l_p^2}, 
\end{eqnarray}
where $S_0=s(t)a(t)^3V_0$ is a conserved quantity representing the radiation entropy per unit co-moving volume. The variable $s(t)$ denotes the entropy density of radiation. To facilitate dimensional analysis, the unit volume $V_0$ is not explicitly set to 1. Accordingly, the special class of spherical systems in the radiation universe is precisely the spheres of radii $R=c_0\,a(t)^3$ with $0<c_0\leq\frac34\frac{k_BV_0}{S_0l_p^2}$.

\section{Holographic bound and a universe composed of black holes and a cosmological constant}

Considering that the critical condition for the holographic bound in static space-times corresponds exactly to black holes, we conjecture that black hole entropy plays a vital role in defining a class of spherical systems that satisfies the above requirement. In light of the conjecture, we study a universe composed of black holes and a cosmological constant. The role of the cosmological constant is solely to control the evolution of the universe. If we can find such a special class of spherical systems in this universe, the result may apply to all cosmological models.

{As a heuristic example, we begin by considering a static universe that is assumed to be homogeneous and isotropic. Here, ``static" refers to a constant scale factor, but black holes are still allowed to merge. For simplicity, the contribution of black hole motion to the mass-energy density of the universe is neglected. Furthermore, energy losses associated with black hole mergers and Hawking radiation are also disregarded. According to the holographic bound, for a spherical system of radius $R$ in this universe, the boundary area should satisfy
\begin{eqnarray}\label{5}
	\frac{ \pi R^2 k_B}{l_p^2}\geq \frac{4\pi G k_B} {\hbar c} {\bigg(\sum_i M_i\bigg)^2}\geq \frac{4\pi G k_B} {\hbar c} \sum_i M_i^2,
\end{eqnarray}
where $M_i$ represents the mass of the $i$-th black hole in the spherical system. The first inequality indicates that even if all black holes in the spherical system collapse into a single black hole, the entropy contained within the spherical system still obeys the holographic bound. Since black holes are evenly distributed in the universe, black hole mergers everywhere in the universe should also be equiprobable. We use the average mass-energy density $\bar\rho$ of the universe to characterize the mass-energy density of the spherical system. Hence, the radius $R$ that satisfies Eq.~(\ref{5}) is given by
\begin{eqnarray}\label{7}
{R\leq R_c= c\left(\frac{8\pi G }{3} \bar \rho\right)^{-\frac12}.}
\end{eqnarray}
Here, $R_c$ can be interpreted as a critical radius, which, consistent with expectations, is a constant for a fixed $\bar \rho$. The critical radius $R_c$ in relation to the spherical system in the static universe is similar to the Schwarzschild radius of a black hole in relation to a spherical system encompassing the black hole in static space-times. It should be noted that $R_c$ is the upper bound on the radius of the spherical system, but the Schwarzschild radius of a black hole is the lower bound on the radius of a system encompassing the black hole.

If the universe evolves, the critical radius $R_c$ should evolve as well. {Here, ``evolve'' implies not only the merging of black holes but also accounts for the time-dependent nature of the scale factor.} To simplify the calculation, we assume that black holes have an average mass of $\bar M$ and use this value to characterize the mass-energy density of the spherical system. Since black holes are continually merging, they can be modeled as a form of dust with a time-dependent average mass $\bar{M}(t)$ and a particle number density $n(t)$. We define an annihilation rate to quantify the frequency of black hole mergers:
\begin{eqnarray}\label{9}
	\Gamma=\frac{\text d N}{\text d t}\frac{1}{N}=\frac{\text d n}{\text d t}\frac{1}{n}+3H,
\end{eqnarray}
where $H$ is the Hubble parameter, $N$ is the number of black holes per unit co-moving volume, and {$n=N/(a^3\,V_0)$}. We define  $n(t_0)=n_0$ as the black hole number density at time $t_0$. Then, the number of black holes per unit co-moving volume at time $t$ can be written as $N(t)=N_0\,\text{exp}\left[\int_{t_0}^{t}\Gamma\,\text d t \right]$,
where {$N_0=N(t_0)=a_0^3\,V_0\,n_0$} is the number of black holes per unit co-moving volume at time $t_0$. According to the preceding assumptions, the total mass-energy of black holes per unit co-moving volume remains constant and is unaffected by black hole mergers. Hence, the average mass of black holes at time $t$ satisfies the relation $\bar M(t) N(t)={\bar M_0\,N_0}$, where $\bar M_0$ is the average mass of black holes at time $t_0$.
 
Now our task shifts to defining a special class of spherical systems, within which the holographic bound maintains validity  at all times, independent of the parameter $\Gamma$. For a spherical system of radius $R$, the number of black holes contained within it can be expressed as $N_s(t)=\frac43\pi {R^3}N(t)/(a^{3}V_0)$. The holographic bound yields $4\, G^2 c^{-4} \bar M(t)^2\,N_s(t)\leq  R^2$. It follows that the radius $R$ need satisfy  the following condition:
\begin{eqnarray}\label{17}
	R\leq R_c=\frac{3\,c^{4}a^3V_0}{16\,\pi\, G^2\bar M_0^2 N_0^2}N_0\, \text{exp}\left[\int_{t_0}^{t}\Gamma\,\text d t \right].
\end{eqnarray}
If there exists a non-vanishing minimum function for the r.h.s. of the inequality and it is independent of $\Gamma$, then the non-vanishing minimum function can  be regarded as a critical radius $R_c$ of the spherical system. Thus, we need determine the lower bound on $N(t)$. If the minimum value of $N(t)$ vanishes, the existence of $R_c$ becomes highly improbable. Otherwise, the spherical system must have a non-vanishing critical radius $R_c$. 

In a universe with an initial singularity, as $a(t)$ approaches zero, the behavior of $N(t)$ is ambiguous. This suggests that the presence of the initial singularity may hinder our definition of the critical radius $R_c$. In this case, we require that the universe composed of black holes and a cosmological constant did not originate from an initial singularity. At the beginning of the universe, we have $N(t=0)>0$. As the universe expands and black holes undergo mergers, $N(t)$ monotonically decreases. We postulate the existence of a critical black hole number density, below which black hole mergers become forbidden. A non-vanishing critical black hole number density ensures a non-vanishing minimum value of $N(t)$. Typically, the critical black hole number density  relies on three factors: the expansion rate of the universe, the relative velocity of black holes, and the average mass of black holes. Therefore, even within this simplified cosmological model, calculating the critical black hole number density remains a complex task. Fortunately, the upper bound imposed by the speed of light on the relative velocity of black holes enables the derivation of a minimum critical black hole number density that is independent of both the relative velocity and average mass of black holes. It is well known that the expansion of the universe can prevent two objects from encountering each other, even if they are moving toward one another at nearly the speed of light. Therefore, when the relative velocity of black holes approaches the speed of light, the critical black hole number density must be minimum across all possible relative velocities. Certainly, under such an extreme condition, the mass of the black hole would also tend towards infinity. However, since the relative velocity of black holes has already reached its maximum possible value, changes in the black hole mass no longer affect our calculation of the critical black hole number density. At this point, the minimum critical black hole number density is solely determined by the expansion rate of the universe. If the minimum critical black hole number density is non-vanishing, it can be used naturally to give a non-vanishing minimum value of $N(t)$. We take an arbitrary black hole as an observer located at the origin. For the black hole closest to the observer, the co-moving coordinate distance between them can be approximated by the average co-moving coordinate distance $\bar r$ among black holes, which increases over time due to black hole mergers. The recessional velocity of the closest black hole is assumed to be approximately linear, i.e.,  $v_{\bar r}=H\,\bar D+v_{p}$, where $\bar D=a(t)\,\bar r$ is the average proper distance and $v_p$ is the peculiar velocity of the closest black hole. If $v_{\bar r}\geq c$, then black holes can no longer encounter each other. In this case, we have $H\,\bar D\geq {c-v_{p}}$. In the limiting case where $v_p\rightarrow-c$, the lower bound on the average proper distance becomes $\bar D=\frac{2c}H$. In the following, we explore the evolution of $N(t)$ in relation to the expansion rate of the universe.

We begin by considering that the universe is decelerating, allowing us to set the cosmological constant to zero. In this case, the scale factor evolves as $a(t)=b\, t^{\frac23}$ with $b$ being a constant. The critical co-moving coordinate distance between the observer and the closest black hole is denoted by $\bar r_c=\frac{2c}{a(t)H}=\frac{3\,c}{b} t_c^{1/3}$, where $t_c$ is a critical time to be determined. When {$\bar r\geq\bar r_c$}, black holes are no longer able to merge. The critical black hole number density at this moment is {$n(t_c)=1/(\frac{4\pi}{3}\bar D^3)=(36\pi c^3\,t_c^3)^{-1}$}. Hence, the number of black holes {per unit co-moving volume} at the critical time is $ N(t_c)=\frac{3\,a^3V_0}{{4\,\pi\bar D^3}}=\frac{3\,V_0}{4\,\pi}\bar r_c^{-3}=\frac{b^3V_0}{36\,\pi c^3 t_c}$, which is the lower bound on $N(t)$. Next, we analyze the value of $\bar r_c$. If $\bar r_c$ is a finite constant, then the lower bound on $N(t)$ is non-vanishing. {Recalling} $v_{\bar r}=H\,\bar D+v_{p}$, the recessional velocity of the closest black hole can be expressed as {$v_{\bar r}=\frac23\,b\,t^{-\frac13}\bar r_c+v_p$. Assuming $\bar r_c$ is finite, it follows that $v_{\bar r}$ decreases over time. As $v_{\bar r}$ decreases,} black holes can continue to merge, which contradicts the assumption that no mergers occur when {$\bar r\geq\bar r_c$}. Therefore, no finite {$\bar r_c$} exists beyond which black holes are guaranteed to remain isolated and the lower bound on $N(t)$ must vanish as time progresses. This suggests that the spherical system may not exist in a decelerating universe composed of black holes.

Now, let us assume that the universe is undergoing accelerated expansion. When the cosmological constant is non-vanishing, the scale factor turns into
\begin{eqnarray}\label{22}
    a(t)=e^{-\sqrt{\frac{\Lambda}{3}}c\,t}\left(\frac12 e^{\sqrt{3\Lambda}\,c\,t}-\frac{4\pi\,G\,\bar\rho_0}{c^2\Lambda}\right)^{\frac23}.
\end{eqnarray}
To ensure the continued acceleration of the expansion, we require $\frac{8\pi\,G\,\bar \rho_0}{\Lambda\,c^2}\leq 2-\sqrt{3}$. Our goal remains to determine a non-vanishing lower bound on $N(t)$.  As in the previous case, the average proper distance $\bar D$ need satisfy the condition $\bar D\,H\geq {c-v_{p}}$. Using the scale factor~(\ref{22}), the critical average co-moving coordinate distance between black holes is given by
	\begin{eqnarray}\label{23}
	\bar r_c=\frac{2^{5/3}\sqrt{3}\,c^2\sqrt{\Lambda}\,e^{\sqrt{\frac{\Lambda}{3}}c\,t_c}\left( e^{\sqrt{3\Lambda}\,c\,t_c}-\frac{8\pi\,G\,\bar\rho_0}{\Lambda\,c^2}\right)^{\frac13}}{c^2{\Lambda}\,e^{\sqrt{3\Lambda}\,c\,t_c}+8\pi\,G\,\bar\rho_0},
\end{eqnarray}
{where $t_c$ is still a critical time to be determined}. Given the constraint $\frac{8\pi\,G\,\bar \rho_0}{\Lambda\,c^2}\leq 2-\sqrt{3}$, it can be shown that $\frac{\text d\bar r_c }{\text d t_c}\leq0$ for all $t_c\geq0$. Then, the maximum value of $\bar r_c$ occurs at $\bar r_c(t_c=0)$, and the lower bound on $N(t)$ is given by $N(t_c=0)>0$. Therefore, the radius $R$ of the spherical system need satisfy
\begin{equation}\label{26}
	\begin{split} 
	R\leq R_c=\frac{\sqrt{3}(c^2\Lambda+8\pi\,G\,\bar\rho_0)^3V_0^2}{2048\,G^2\sqrt{\Lambda}\,\pi^2\bar M_0^2 N_0^2(c^2\Lambda-8\pi\,G\,\bar\rho_0)}a^3=f_0a^{3},
\end{split}
\end{equation}
where $f_0$ is a non-vanishing constant with the dimension of length. Hence, in an accelerating universe composed of black holes and a cosmological constant, there exists a special class of spherical systems (with radius $R\leq R_c= f_0a^{3}$), within which the entropy of black holes consistently adheres to the holographic bound. 

In fact, it can be demonstrated that the fundamental condition for the feasibility of the definition~(\ref{26}) is the existence of an upper bound on the entropy per unit co-moving volume, regardless of the expansion rate of the universe and the presence of an initial singularity. For a spherical system defined by~(\ref{26}), the entropy contained within the system satisfies 
\begin{equation}\label{27}
	\begin{split} 
\frac{\frac{{4\,\pi}}3 (f_0a^{3})^3}{a^3V_0} S(t)\leq\frac{\frac{{4\,\pi}}3 (f_0a^{3})^3}{a^3V_0} S_{max}\leq\frac{k_B  (f_0a^{3})^2}{l_p^2},
	\end{split}
\end{equation}
where $S(t)$ is the entropy per unit co-moving volume at time $t$, and $S_{mas}$ is the maximum value of $S(t)$ for the entire evolution of the universe. Then, we have 
\begin{equation}\label{28}
	\begin{split} 
		0<f_0\leq \frac{3\,k_B V_0}{16\,\pi\,l_p^2\, S_{max}}.
	\end{split}
\end{equation}
Therefore, we can conclude that for any cosmological model, as long as the entropy per unit co-moving volume (or even any volume) has an upper bound, there exists a critical radius $R_c = f_0 a^{3}$ with $f_0$ being a non-vanishing constant. When the radius of a spherical system satisfies $R\leq R_c$, the entropy contained within the system can consistently adhere to the holographic bound. If we can prove that the entropy per unit co-moving volume has an upper bound for all cosmological models, we can assert that the definition~(\ref{26}) is a universal result. However, this poses a significant challenge, as in a decelerating universe originating from an initial singularity, the entropy per unit co-moving volume may become infinitely large in both the early and late stages of the universe.

\section{Holographic bound and a general cosmological model}

In the previous part, we have defined a special class of spherical systems in cosmology, within which the entropy of matter constantly adheres to the holographic bound, based on the assumption that the entropy per unit co-moving volume is bounded from above in the universe. In this part, we further explore the holographic bound in a general cosmological model and aim to establish a more universal definition from the perspective of the entropy per unit co-moving volume.

We start with an arbitrary cosmological model, in which the entropy of matter per unit co-moving volume is described by a time-dependent function $\sigma(t)$. The holographic bound requires the entropy of matter contained within a spherical system of radius $R$ to satisfy
\begin{eqnarray}\label{32}
    S=\frac{\frac43 \pi R^3}{a(t)^3V_0}\cdot \sigma(t)\leq \frac{k_B\,\pi R^2}{l_p^2}.
\end{eqnarray}
Given a mass-energy density $\rho(a)$, the corresponding entropy function $\sigma(t)$ may depend on multiple factors, and whether it possesses a universal upper bound is unknown. However, for any $\rho(a)$, we can define an auxiliary entropy function $\sigma_x(t)$ such that $\sigma_x(t)\geq \sigma(t)$ at all times, with an additional constraint that $\sigma_x(t)$ depends solely on the scale factor. If such a function $\sigma_x(t)$ exists, it can be used to derive a  general form of the critical radius $R_c$ applicable to any cosmological model. In the following, we present a method for constructing such a function $\sigma_x(t)$.

When all the matter contained within a unit co-moving volume is treated as a black hole of the same mass-energy $m(a)=a^3V_0\rho(a)$, the entropy of this black hole can be expressed as
	\begin{eqnarray}\label{33}
\sigma_{BH}(t)=\frac{4\pi Gk_B}{\hbar c} a^6V_0^2\rho(a)^2.
\end{eqnarray}
If $\sigma_{BH}(t)$ is the maximum entropy within unit co-moving volumes that share the same mass-energy, and if we can define a class of spherical systems based on $\sigma_{BH}(t)$ such  that the entropy contained within these systems adheres to the holographic bound, then this definition should be applicable to other universes as well. According to the Bekenstein-Hawking entropy formula and the generalized second law of thermodynamics, the inequality $\sigma_{BH}(t) \geq \sigma(t)$ ``appears to'' hold for any form of matter.

To analyze the evolution of $\sigma_{BH}(t)$, we employ the Friedmann equation. Assuming the universe contains only a single component of matter, characterized by an energy density $\rho(a)$, one can use the Friedmann equation to derive that $\sigma_{BH}(t)\sim H^4a^6$. Therefore, the radii of these special spherical systems need satisfy 
\begin{eqnarray}\label{34}
\!\!\!\!\!\!\!	R\leq R_c=\frac{k_B\,\pi}{l_p^2}\frac{\hbar c}{4\pi Gk_B} \frac{16\pi G^2}{3V_0} H^{-4} a^{-3}=\frac{4\pi}{3V_0} \frac{c^4} {H^{4}a^3}.
\end{eqnarray}
The definition of $R_c$ is independent of the specific components and expansion of the universe. As such, it provides a universal definition for  these spherical systems ensuring that the entropy contained within them adheres to the holographic bound. Strictly speaking, this result relies only on two assumptions: 1) The evolution of the universe {is governed by} the Friedmann equation; 2) Among unit co-moving volumes with the same mass-energy, when the mass-energy is converted into a black hole, the entropy reaches its maximum. We can regard $\frac{4\pi}{3V_0}\frac{c^4} {H^{4}a^3}$ as a characteristic length scale in the universe. It implies that as long as the radius of any spherical system in the universe does not exceed this scale, the entropy contained within it can remain compliant with the holographic bound. Defining $R=g_0 H^{-4}a^{-3}$ with $g_0\leq \frac{4\pi}{3V_0}c^4$, and differentiating it with respect to time, we have $\frac{\text d R}{\text d t}=g_0\frac{\dot a^2-4a\,\ddot a}{\dot a ^5}$. If the universe is undergoing decelerated expansion, $R$ increases over time. But, in the case of accelerated expansion and $\ddot a>\frac{\dot a^2}{4a}$, $R$ decreases over time. 

It is important to emphasize that the condition $\sigma_{BH}(t)>\sigma(t)$ actually does not hold in all cosmological models. For example, in an expanding universe consisting solely of photons, $\sigma(t)$ remains constant while the internal energy per unit co-moving volume decreases with the scale factor.  As a result, for sufficiently large scale factors, one finds that $\sigma_{BH}(t)<\sigma(t)$. Hence, the definition (\ref{34}) is not the most universal either. Regarding this counterexample, we provide an analysis at the end. 

\section{Discussion} 

We have found two specific definitions for spherical systems in cosmology that ensure the entropy contained within them adheres to the holographic bound throughout the evolution of the universe, regardless of the universe's components. However, both definitions exhibit notable limitations, as we have discovered respective counterexamples. The definition (\ref{26}) is strictly valid only for cosmological models where the entropy per unit co-moving volume is bounded from above, while the definition (\ref{34}) fails in a radiation universe.

For the definition (\ref{26}), the requirement that the entropy per unit co-moving volume has an upper bound, is almost certain to fail near the singularity of the universe. If a cosmological model satisfies this requirement, compliance with other entropy bounds can be achieved by appropriately adjusting the exponent of the scale factor in the definition (\ref{26}). For instance, with regard to the Bekenstein bound, the radii of spherical systems are given by $R\leq R_c=f_0 a$ with $f_0$ being a non-vanishing constant. In this case, counterexamples to the holographic bound may also arise for other entropy bounds, not due to a flaw in the entropy bounds themselves, but rather because of the similar definitions of spherical systems.

The fundamental reason why definition (\ref{34}) fails in a radiation universe is that converting the mass-energy of photons within a system into a black hole may reduce the total entropy of the system, which does not occur in other known ordinary matter. This counterexample suggests that, in order to properly apply the holographic bound (as well as other entropy bounds) in cosmology, we must first clarify the maximal entropy state among all systems with the same mass-energy. If this problem can be resolved, it may pave the way toward identifying the most universal definition of such a special class of spherical systems. It now appears that the common assumption that black holes always represent the maximal entropy state is not entirely accurate.

\vspace{10pt}

\noindent {\bf Acknowledgments}

\noindent This work was supported by the National Natural Science Foundation of China (Grants No. 11873001, No. 12047564, and No. 12247142), and the China Postdoctoral Science Foundation (Grant No. 2024M753825).

\end{document}